\documentclass[osajnl,twocolumn,preprint,showpacs,superscriptaddress,floatfix,10pt]{revtex4-1} 
\usepackage{graphicx}
\usepackage{color}
\usepackage{amsmath,amsfonts,amssymb}
\def\0#1{\xout{#1}}
\def\1#1{{\color{black}#1}}
\def\2#1{{\color{blue}#1}}
\def\3#1{{\color{magenta}#1}}
\def\4#1{{\color{yellow}#1}}

\newcommand{\de}{\,\mathrm d}
\begin{document}

\title{Phase retrieval via regularization in self-diffraction based spectral interferometry}

\author{Simon Birkholz}\email{Corresponding author: birkholz@mbi-berlin.de}
\affiliation{Max-Born-Institut, Max-Born-Stra\ss e 2a, 12489 Berlin, Germany}

\author{G\"unter Steinmeyer}
\affiliation{Max-Born-Institut, Max-Born-Stra\ss e 2a, 12489 Berlin, Germany}

\author{Sebastian Koke}
\affiliation{Max-Born-Institut, Max-Born-Stra\ss e 2a, 12489 Berlin, Germany}
\affiliation{present address: Coherent Laser Systems GmbH \& Co.~KG, Garbsener Landstr.~10, 30419 Hannover, Germany}

\author{Daniel Gerth}
\affiliation{Johannes-Kepler-Universit\"at, Doctoral Program Computational Mathematics, Altenbergerstra\ss e 69, 4040 Linz, Austria}

\author{Steven B\"urger}
\affiliation{Technische Universit\"at Chemnitz, Fakult\"at f\"ur Mathematik,  09107 Chemnitz, Germany}

\author{Bernd Hofmann}
\affiliation{Technische Universit\"at Chemnitz, Fakult\"at f\"ur Mathematik,  09107 Chemnitz, Germany}

\date{\today}

\begin{abstract}\vspace{1em}\footnotesize{
\noindent A novel variant of spectral phase interferometry for direct electric-field reconstruction (SPIDER) is introduced and experimentally demonstrated. Other than most previously demonstrated variants of SPIDER, our method is based on a third-order nonlinear optical effect, namely self-diffraction, rather than the second-order effect of sum-frequency generation. On one hand, self-diffraction (SD) substantially simplifies phase-matching capabilities for multi-octave spectra that cannot be hosted by second-order processes, given manufacturing limitations of crystal lengths in the few-micrometer range. On the other hand, however, SD SPIDER imposes an additional constraint as it effectively measures the spectral phase of a self-convolved spectrum rather than immediately measuring the fundamental phase. Reconstruction of the latter from the measured phase and the spectral amplitude of the fundamental turns out to be an ill-posed problem, which we address by a regularization approach. We discuss the numerical implementation in detail and apply it to measured data from a Ti:sapphire amplifier system. Our experimental demonstration used 40-fs pulses and a 500\,$\mu$m thick BaF${}_2$ crystal to show that the SD SPIDER signal is sufficiently strong to be separable from stray light. Extrapolating these measurements to the thinnest conceivable nonlinear media, we predict that bandwidths well above two optical octaves can be measured by a suitably adapted SD SPIDER apparatus, enabling the direct characterization of pulses down to single-femtosecond pulse durations. Such characteristics appear out of range for any currently established pulse measurement technique.}
\end{abstract}


\maketitle

\section{Introduction}

\noindent Ultrashort light pulses can nowadays be generated in a large spectral range, with pulse durations reaching the single-cycle regime \cite{Goulielmakis,Kaertner,Baltuska}. At 800\,nm central wavelength, the intensity full width at half maximum of a single-cycle pulse encompasses only 2.7\,fs, and the spectrum of such short pulses covers more than an optical octave, i.~e., a wavelength ratio of more than $1$:$2$. At such extreme parameter ranges, the slowly-varying envelope approximation fails, and dependable pulse characterization becomes extremely difficult. Traditionally, characterization of few-cycle optical pulses employs interferometric autocorrelation \cite{Diels} using second-harmonic generation (SHG) in a nonlinear optical crystal. This $\chi^{(2)}$ based method inevitably fails when the bandwidth exceeds the optical octave and when portions of the fundamental start to interfere with the second harmonic. While this problem can be alleviated by resorting to a non-collinear geometry, beam smearing \cite{
geosmear} may then give rise to a loss of temporal resolution in this geometry. Similar considerations can also be made for $\chi^{(2)}$ based frequency-resolved optical gating (FROG) variants \cite{FROG}. Collinear implementations of FROG \cite{collinearFROG,IFROG1,IFROG2} are not suitable for spectra beyond the optical octave whereas non-collinear implementations may suffer from beam smearing. While beam smearing may be overcome for pulses that are about two cycles long \cite{BaltuskaGS}, this problem appears virtually impossible to solve for single-cycle pulses \cite{notegeosmear,Baltuska}.

\begin{figure}[tbh]
\includegraphics[width=8cm]{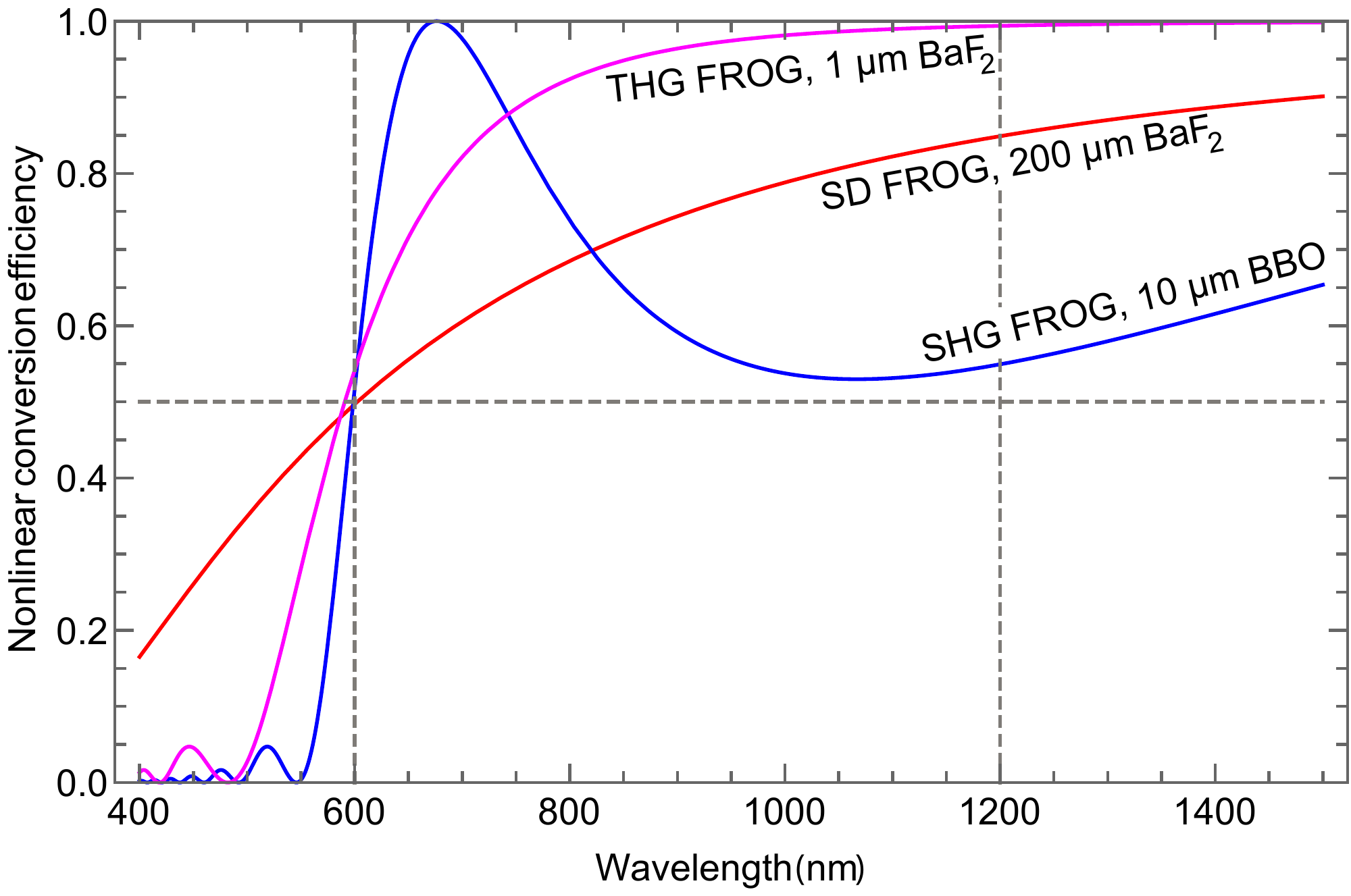}
\vspace{-1em}
\footnotesize{\caption{\label{fig:pmFROG} Spectral dependence of phase matching efficiency $\propto {\rm sinc}(\Delta k L / 2)$ of degenerate pulse characterization techniques including FROG, d-scan, MIIPS, and autocorrelation. $\Delta k$ is the wavevector mismatch. Parameters have been optimized to yield $>50\%$ nonlinear conversion efficiency in the range from 600 to 1200\,nm indicated by dashed lines. Blue curve: standard SHG FROG. $L=10\,\mu$m thick BBO crystal cut at $\vartheta=35^\circ$ for type-I phase matching. Magenta curve: THG FROG in $L=1\,\mu$m barium fluoride. Red curve: SDFROG employing $L=200\,\mu$m barium fluoride. External beam crossing angle $\alpha=5^\circ$.}}
\vspace{-1em}
\end{figure}

This fundamental dilemma of correlation-based methods can also be extended to other methods like multiphoton intrapulse interference phase scan (MIIPS) \cite{MIIPS} and dispersion scan \cite{dscan} as long as they rely on second-order nonlinear processes. As both methods depend on a collinear geometry, they can ultimately not separate the fundamental and the second harmonic of octave-spanning spectra. Of course, one can resort to third-harmonic generation \cite{Nagy} instead, but then has to suffer from the much tighter phase-matching constraints of this process, see Fig.~\ref{fig:pmFROG}. Unfortunately, the more favorable four-wave mixing variants of the third-order nonlinearity, namely self-diffraction \cite{SDFROG} and transient grating \cite{TGFROG}, cannot be used in collinear geometries. These considerations make it clear that pulse characterization of near-infrared pulses faces severe problems when the pulse duration approaches a single optical cycle or when spectra extend beyond one optical octave.

\begin{figure}[tb]
\includegraphics[width=8cm]{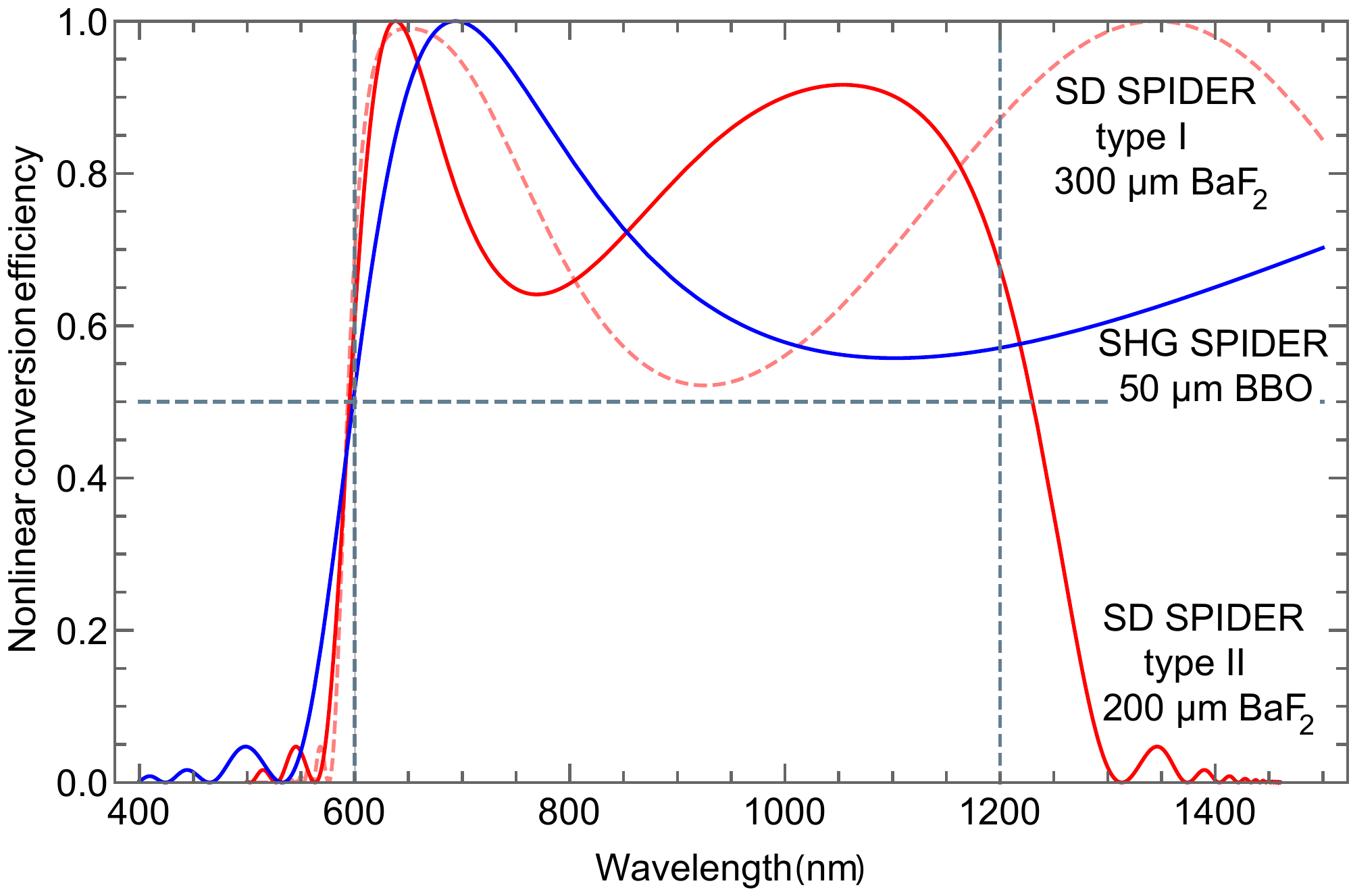}
\vspace{-1em}\footnotesize{\caption{\label{fig:pmSPIDER} Similar as in Fig.~\ref{fig:pmFROG}, but for non-degenerate characterization techniques like SPIDER, cf.~Fig.~\ref{fig:scheme} for the various phase matching schemes. Wavelength refers to the replica pulse. Ancilla was assumed as monochromatic. Blue curve: standard SPIDER based on sum-frequency generation in a 50\,$\mu$m thick BBO crystal cut at $\vartheta=44^\circ$ for type-II phase matching. An 800\,nm ancilla was assumed in the extraordinary (fast) axis of the crystal. Red curves: SD SPIDER in BaF${}_2$ at an external beam crossing angle $\alpha=5^\circ$. Solid curve: interaction of two 1030\,nm ancilla photons with one broadband replica photon (type I). Dashed curve: interaction of two replica photons with one 750 nm ancilla photon (type II). Insets show phase matching variants for SD SPIDER.}}\vspace{-1em}
\end{figure}

Interferometric techniques appear to offer an alternative here and have been used for characterization of some of the shortest optical pulses generated to date \cite{Gallmann,MSPIDER,SEASPIDER,comparison,2DSI,2DSI2,Yamashita2DSI,Baltuska}. The prototypical method is spectral-phase interferometry for direct electric-field characterization \cite{SPIDER,SPIDER2,SPIDER3} (SPIDER). This method relies on $\chi^{(2)}$ based upconversion of two delayed replicas of the pulse under test with a third chirped pulse. This so-called ancilla pulse is often derived from the laser under test, which makes the SPIDER method self-referenced. Alternatively, some implementations use an external coherent reference pulse \cite{MSPIDER,Yamashita2DSI}. The upconverted replica pulses are then spectrally resolved in a spectrometer. SPIDER analytically retrieves the spectral phase of the measured pulse from the spacing of the fringes in the resulting interferogram. More precisely, the phase information is encoded in deviations from an equal
frequency spacing of the observed fringe pattern. The $\chi^{(2)}$ based mixing process can be implemented in a non-collinear geometry without compromising temporal resolution. While geometrical smearing, in principle, also affects the SPIDER method, this artifact only reduces the obtainable fringe contrast but not the fringe spacing. This fact makes interferometric techniques widely immune against beam smearing. Moreover, the SPIDER method is relatively insensitive towards phase-matching constraints as the method does not rely on a spectrally flat conversion efficiency of the $\chi^{(2)}$ process. As shown in Fig.~\ref{fig:pmSPIDER}, one can exploit the favorable properties of type-II phase matching and obtain already an octave-spanning conversion efficiency with a relatively thick 50\,$\mu$m BBO crystal. For the case of BBO, the narrowband ancilla beam has to be placed in the fast crystal axis. Going to the extremes, a nonlinear conversion bandwidth of up to two octaves can be obtained by careful
optimization of the crossing angle \cite{Yamashita2DSI}. In comparison, type-I phase-matching already requires 10\,$\mu$m thin crystals to obtain octave conversion bandwidth in the wavelength degenerate case of non-interferometric techniques (Fig.~\ref{fig:pmFROG}). This minimum thickness clearly marks a limit for all these approaches. Current endeavors to coherently combine the entire wavelength range from 450\,nm to 2.4\,$\mu$m \cite{Optica} would require crystal thicknesses of $3\,\mu$m for the SPIDER measurement and $1\,\mu$m for the reference measurement, see subsequent discussion. The sub-cycle pulse durations of 1.9\,fs predicted in \cite{Optica} can clearly not be hosted by any available pulse characterization technique to date.

At pulse durations in the single-cycle regime, SPIDER therefore seems to have a clear advantage over other techniques. Nevertheless, SPIDER often critically relies on a reference measurement. To this end, one typically uses the same replica pulses as in the SPIDER measurement, but converts them with the degenerate SHG process \cite{RevSciInstrum}. Unfortunately, this conversion process underlies the same restrictions as those seen by FROG and other non-interferometric techniques, cf.~Fig.~\ref{fig:pmFROG}. Here we now discuss how to use a $\chi^{(3)}$ based self-refraction process for implementing SPIDER. As input and output wavelength range coincide in all of the four-wave mixing variants, one does not require any nonlinear conversion for the calibration measurement but can simply use the replica pair directly for this purpose. Moreover, as indicated by the highly favorable phase-matching properties of the self-diffraction \cite{selfdiffraction} process for FROG, a $\chi^{(3)}$-based SPIDER offers virtually
unlimited bandwidth. Assuming BaF${}_2$ as the $\chi^{(3)}$ material \cite{BaF2}, this advantage is demonstrated in Fig.~\ref{fig:pmFROG}. Here a $200\,\mu$m material path length was assumed, which is still an order of magnitude away from the thinnest feasible thickness of suitable dielectric materials. In fact, self-refraction based SPIDER schemes have been already demonstrated \cite{CLEO,SPIDERonchip,TGSRSI,Kobayashi}, yet did not find widespread use so far. As we discuss below, there are in fact two conceptually different ways how to implement a SPIDER with a four-wave mixing nonlinearity. As four-wave mixing requires three input waves to generate the fourth one, one can either mix two ancillas with the replica or, vice versa, use only one ancilla with two identical replica beams. Both methods have their distinct advantages and disadvantages as will be discussed below.
\begin{figure}[tb]
\includegraphics[width=\columnwidth]{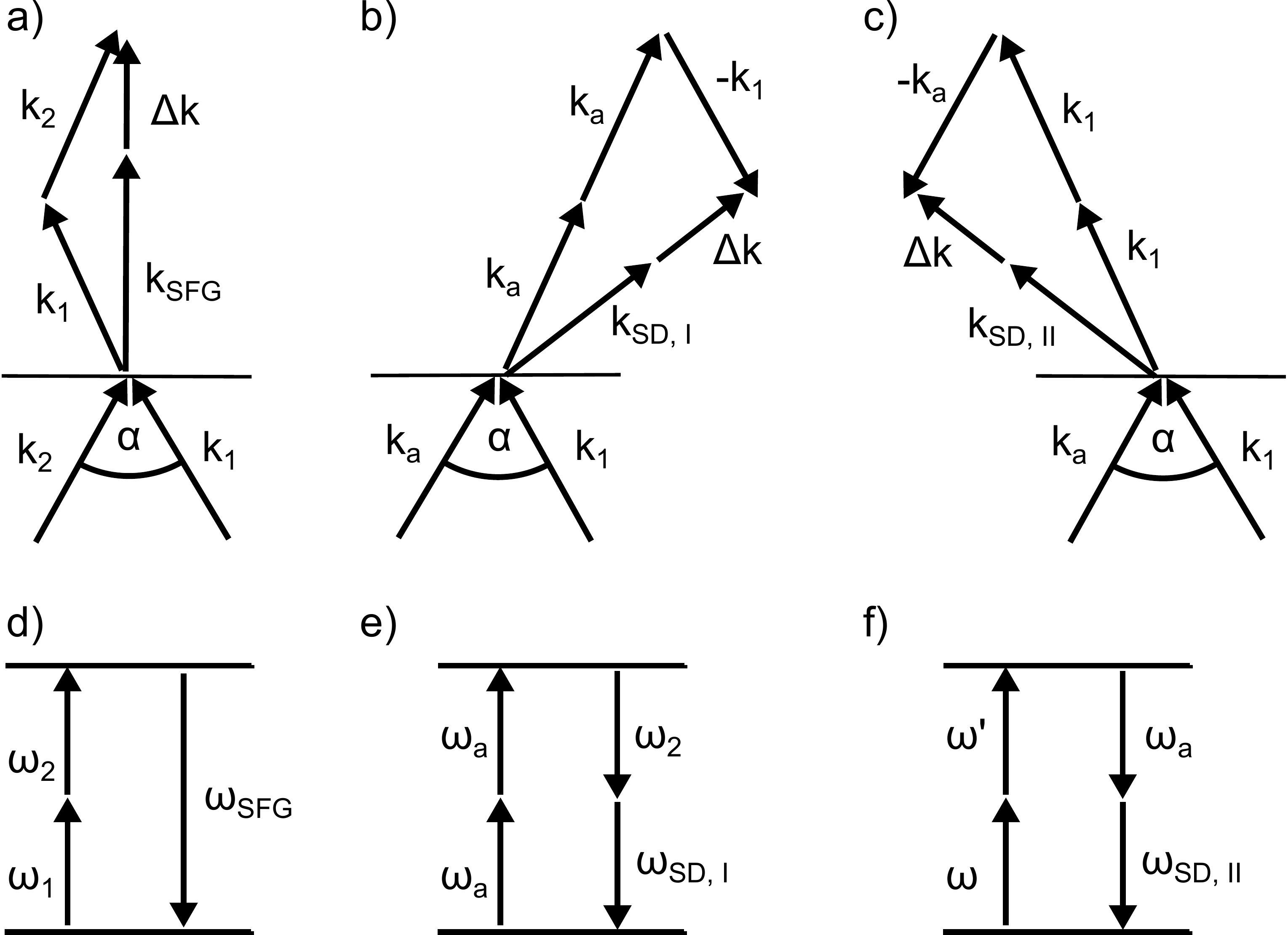}
\vspace{-1em}\footnotesize{\caption{\label{fig:scheme}
Phase matching (a-c) and energy conservation (d-f) in the two possible self-diffraction based SPIDER variants in comparison to conventional SPIDER. (a,d) Conventional SPIDER based on sum-frequency generation. (b,e) Type-I SD SPIDER, with two ancilla photons. (c,f) Type-II SD SPIDER with only one ancilla photon.}}\vspace{-1em}
\end{figure}
\vspace{-1em}
\section{Principle of the SPIDER method}
\label{sec:principles}

\begin{figure}
\includegraphics[width=8cm]{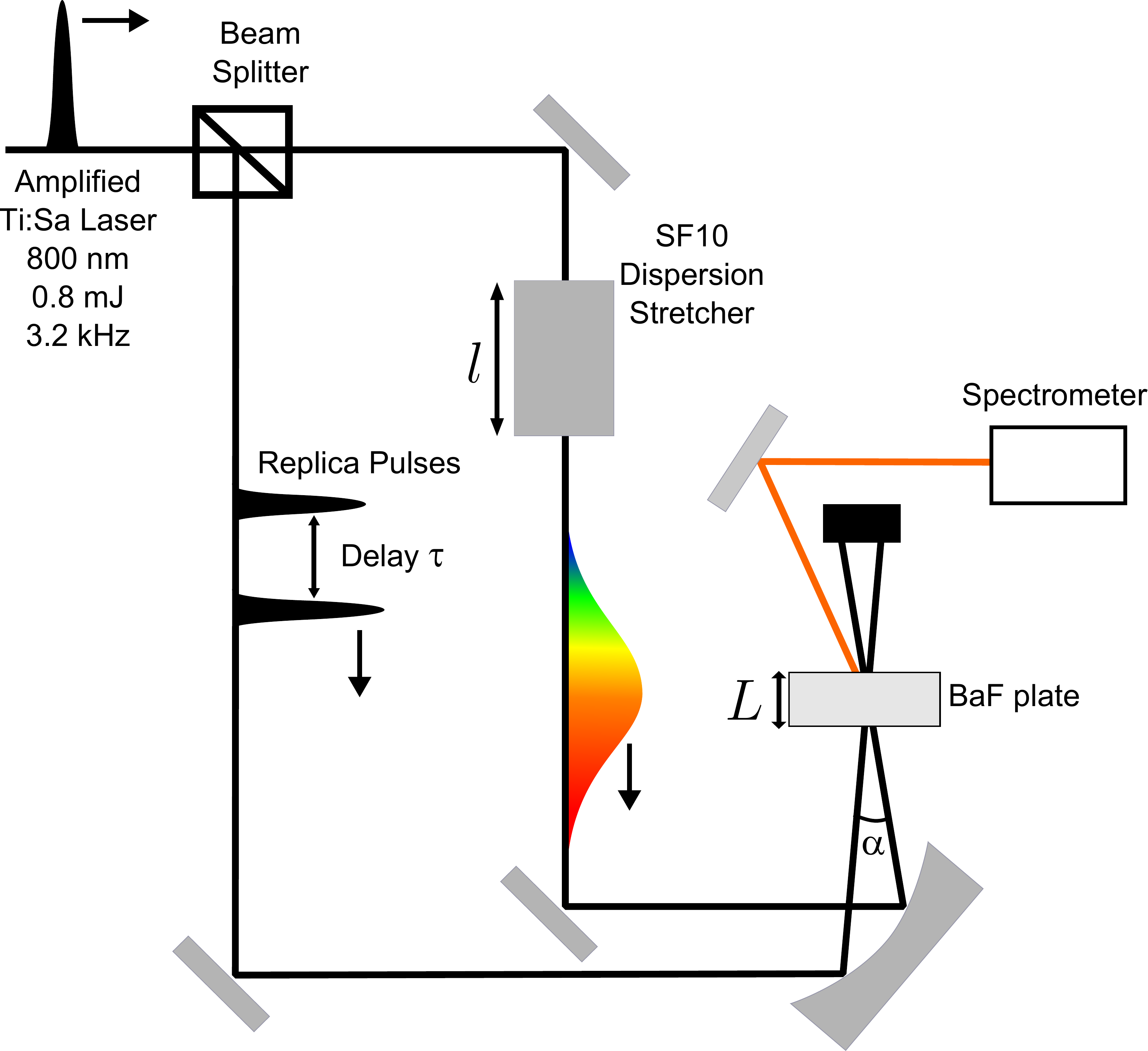}
\caption{\label{fig:setup} Setup. Laser: Ti:sapphire CPA system, delivering pulses up to 800 $\mu$J at a repetition rate of 3.2\,kHz. Pulse duration: 50\,fs. Beam splitter: 175\,$\pm$ 25\,$\mu$m thick glass \'{e}talon. Dispersion stretcher: $l=10\,$cm SF10 glass block. BaF${}_2$ plate: $\chi^{(3)}$ medium, thickness $L=500\,\mu$m. Spectrometer: 0.5\,m grating spectrograph. Line camera: 2048 pixels with a size of 13 $\times$ 500\,$\mu$m${}^2$. Focusing employed an $f=15\,$cm concave mirror in a nearly stigmatic geometry, resulting in an estimated beam radius $w_0\approx 230\mu$m.}
\end{figure}

SPIDER is based on spectral interferometry
\cite{Lepetit}, a method that is used to measure the relative phase between two optical pulses, see Fig.~\ref{fig:setup} for the optical layout of a SPIDER apparatus. As both pulses have to be coherent and have to spectrally overlap, one normally uses this method to measure the effect on the group delay dispersion of one the pulses. Without having a well characterized reference pulse available, spectral interferometry does not allow to measure the spectral phase of an unknown pulse. If two identical pulses are used at a delay $\tau$, an equidistant
spectral interference pattern with a period $\delta\omega = 2 \pi / \tau$ is measured in a spectrograph. Phase differences $\varphi(\omega)$ result in non-equidistant fringe patterns.
Using a Fourier filtering approach
\begin{eqnarray}
\varphi(\omega) = {\rm arg}\bigg[ & & \int\limits_{\tau/2}^{3 \tau/2}
\int\limits_{0}^{\infty}S(\omega')\exp(-i \omega' t') {\rm d}\omega' \nonumber
\\ & & \times \exp(i \omega t') {\rm d}t' \bigg] \label{eq:takeda}
\end{eqnarray}
one can then reconstruct $\varphi(\omega)$ from the measured fringe pattern \cite{Takeda}. The argument function is defined as the imaginary part of the complex logarithm $\arg(x) := {\rm Im} \log(x)$. While spectral interferometry can only be used to measure the relative phase between two pulses, the induction of a spectral offset $\Omega$ between two otherwise identical replicas of a pulse may be utilized to measure the phase of an unknown pulse in a completely self-referenced way \cite{SPIDER,SPIDER2}. The introduction of a spectral shear is the basis of the SPIDER method. In SPIDER, division of the phase difference by the shear $\Omega$ then provides the group delay
\begin{equation}
{\rm GD}(\omega_c)= \left. \frac{\partial \varphi}{\partial \omega}
\right|_{\omega_c} \approx {\varphi(\omega_c+\Omega)-\varphi(\omega_c) \over
\Omega} \label{eq:SPIDERmaster}
\end{equation}
of the unknown pulse as a function of frequency $\omega_c$.


In most previous demonstrations of SPIDER, the shear is generated by sum-frequency generation (SFG).
This method requires a third pulse that can also be deduced from the pulse under test, cf.~Fig.~\ref{fig:setup}.
This ancilla pulse is generated by sending a third replica of the pulse through a dispersive
delay line, e.g., a glass block with length $L$, in which the pulse becomes strongly chirped.
As a result, the carrier frequency $\omega_0(t)$ of the pulse is monotonically increasing with $t$,
and each of the two replica pulse at delay $\tau$ is frequency-shifted by a different amount
in the SFG process. The shear $\Omega$ relates to the group-delay dispersion of the
glass block $\beta_2 l$ via
\begin{equation}
\Omega=\omega_0(t+\tau)-\omega_0(t)=\frac{\tau}{\beta_2 l}. \label{eq:shear}
\end{equation}
Measuring the SPIDER interferogram of the sheared pulses and processing the measured data with Eqs.~(\ref{eq:takeda})-(\ref{eq:shear}) then gives completely self-consistent access to measurement of the spectral phase $\varphi(\omega)$ of a short pulse. Compared to other characterization techniques, SPIDER is relatively immune to spectral efficiency variations of the nonlinear optical process as all relevant information is encoded in the spacing of the fringes, not in their amplitude. Nevertheless, deviations from equidistance are typically very small. Therefore, SPIDER critically depends on an accurate calibration of the spectrograph wavelength scale. In practice, the most dependable method for this calibration is measurement of the fringe pattern of two identical pulses at delay $\tau$. When using the SFG process, this calibration pulse pair can be readily generated via simple second-harmonic generation (SHG) of the same two replicas that are used in SPIDER. Processing this calibration data in the
same manner as the SPIDER data sets, one obtains a phase correction $\epsilon_\varphi(\omega)$, which then needs to subtracted from every subsequent phase measurement. As already discussed, despite its simplicity, the calibration measurement may pose a greater challenge than the SPIDER measurement itself.
\vspace{-1em}
\section{Self-diffraction SPIDER}
\label{sdspider}

Figures \ref{fig:scheme}(b,c,e,f) schematically show the phase matching and energy conservation of four-wave mixing type SPIDER in comparison to regular SPIDER [Fig.~\ref{fig:scheme}(a,d)]. We have chosen the proven self-diffraction (SD) geometry that has been very successfully employed for FROG measurements in the ultraviolet \cite{Graf,SDFROG}. Compared to thin $\chi^{(2)}$ optical crystals, the SD process offers a vast bandwidth that can easily host one octave at 800\,nm. Moroeover, the process is scalable deep into the ultraviolet. In SD FROG, the nonlinear process is never exactly phase-matched, cf.~Fig.~\ref{fig:pmFROG}, i.e., the conversion efficiency contribution ${\rm sinc} ( \Delta k L / 2 )$ from phase matching never reaches unity. The same effect is also seen for SD SPIDER in the immediate vicinity of the ancilla wavelength. For example, the latter was chosen as 800\,nm for the red solid curve in Fig.~\ref{fig:pmSPIDER}, and this central wavelength also exhibits a local conversion efficiency minimum. Phase matching conditions become more favorable at 1100\,nm and even reach unity near 650\,nm for this example case. The most favorable phase-matching conditions can be obtained for mixing of two ancilla photons with broadband replica photons (dashed curve in Fig.~\ref{fig:pmSPIDER}, ancilla at 950\,nm). In the following, we will refer to this process [phase matching scheme Fig.~\ref{fig:scheme}(b)] as type-I SD SPIDER. Slightly inferior performance is expected for type-II SD SPIDER that mixes two photons from the broadband replica with a single photon from the ancilla, cf.~the phase matching scheme in Fig.~\ref{fig:scheme}(c). On the other hand, compared to SD FROG (Fig.~\ref{fig:pmFROG}), a similar material thickness is tolerable in both variants of SD SPIDER to cover the same octave spanning wavelength interval. In contrast, SD FROG is never perfectly phase-matched and only exhibits a short-wavelength cut-off.

While both variants have similar phase matching capabilities, there is a substantial difference in the expected conversion efficiency. In type-I SPIDER, two narrowband electric fields $E_{\rm a}$ mix with one broadband field $E_{\rm r}$, yielding an output signal with power $\propto \left| E_{\rm a}^2 E_{\rm r}^\ast \right|^2$. This process reaches optimum conversion efficiency if the pulse energy of the individual replica pulses as well as the ancilla are chosen identical, similar to regular SPIDER \cite{RevSciInstrum}. If the roles of the beams are reverted, the power follows $\propto \left| E_{\rm r}^2 E_{\rm a}^\ast \right|^2$. Now optimum conversion is obtained if the energy of the ancilla is chosen four times the energy of each replica pulse. As the ancilla beam is temporally stretched by a factor 100 or more compared to the input pulse duration, a correspondingly lower maximum conversion efficiency of the type-I process results. While, in principle, this can be compensated by simply increasing the total input energy by a factor $\approx 4.5$, such energy scaling is ultimately limited by the damage threshold of the material. Nevertheless, it becomes clear that the type-II variant has to be strongly preferred in terms of conversion efficiency.

\begin{figure}[tb]
\includegraphics[width=8cm]{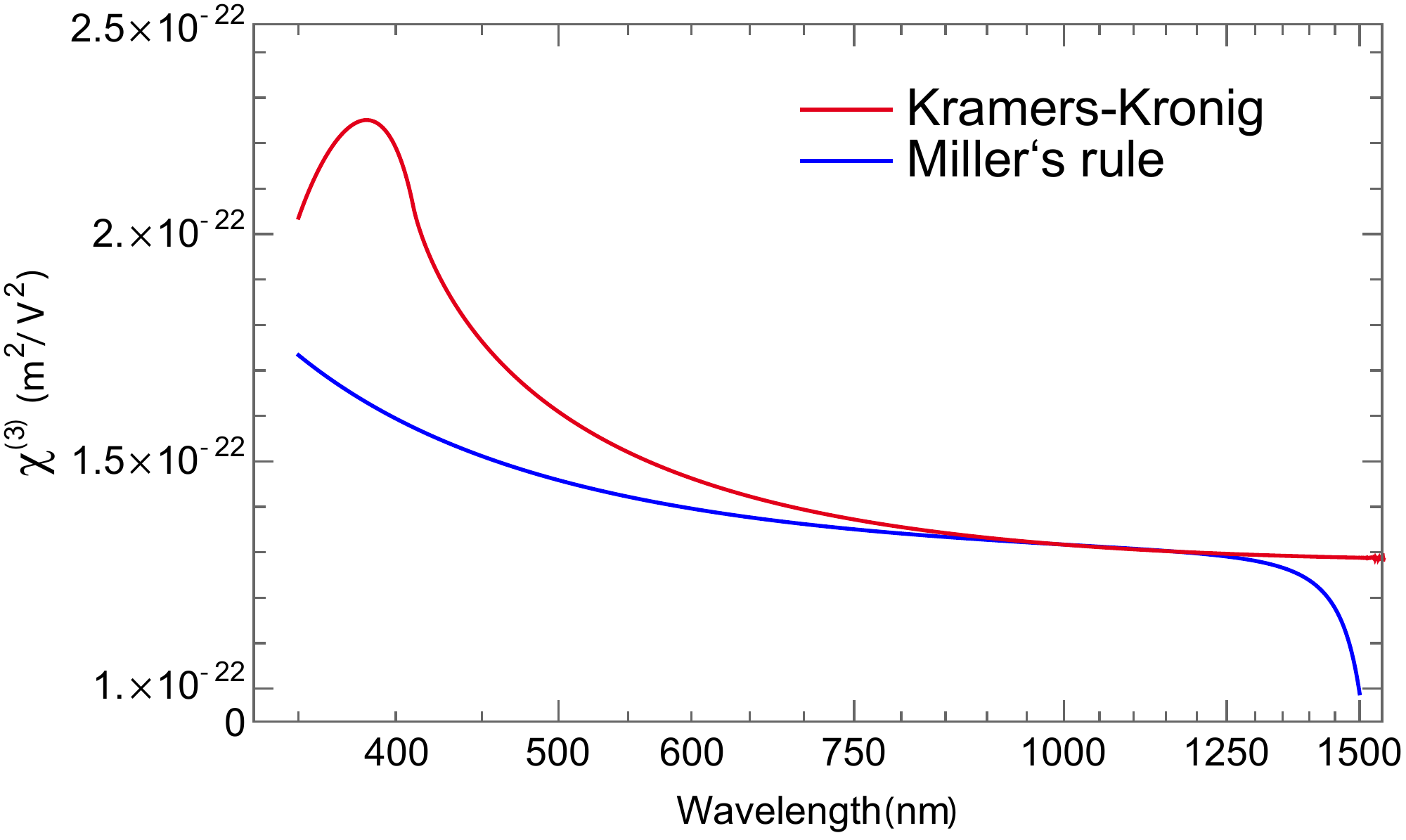}\vspace{-1em}
\footnotesize{\caption{\label{fig:miller}Dispersion of the third-order susceptibility $\chi_3(\omega,\omega,-\omega_{\rm a})$ (singly degenerate case). Ancilla wavelength: 800\,nm. Blue curve: as predicted by Miller's rule, Eq.~(\ref{eq:miller}), using $d=2.79\,$\AA, which has been chosen to obtain a match with measured $n_2$ data at 1064\,nm (cf.~Table II in \cite{Sheik-Bahae2}). Red curve: Computation based on Kramers-Kronig relations according to Eq.~(\ref{eq:SheikBahae}) and \cite{Sheik-Bahae}. The bandgap $E_{\rm g}=9.0\,$eV has been chosen according to Table I in \cite{Sheik-Bahae2}. The refractive index of BaF${}_2$ was modeled according to \cite{nBaF2}. For comparison: For fused silica, a totally degenerate $\chi_3(\omega,\omega,\omega)$ of $(2.0 \pm 0.2) \times 10^{-22} {\rm m}^2/{\rm V}^2$ was reported \cite{Silica}.\vspace{-1em}
}}
\end{figure}

Comparing the role of the shear in SD SPIDER, one can understand from the energy conservation diagrams in Fig.~\ref{fig:scheme}(e,f) that the conjugate field of the ancilla enters in the type II variant whereas the ancilla field enters twice in type I. In the former case, this induces a frequency shift in the opposite direction than in regular SPIDER, i.e., an up-chirped ancilla beam in regular SPIDER has the same qualitative effect as a down-chirped ancilla in SD SPIDER. Moreover, in the type I variant, the obtained frequency shifts are doubled compared to regular SPIDER, but the sign of the shift remains identical. These differences can easily be accomodated by minor adjustments in the phase retrieval. Type-II SD SPIDER bears another slightly hidden problem, namely, it does not measure the phase of the pulse under test but effectively that of its second harmonic. This problem is best understood by analyzing the role of the three interacting waves in the self-diffraction process, giving rise to an output field
\begin{eqnarray}
E_2(\omega) & \propto &  \frac{\omega}{n(\omega)} \int\limits_0^L P_{\rm I,II} (\omega) \exp(-i \Delta k z) {\rm d}z \\
& = &  \frac{\omega}{n(\omega)} P_{\rm I,II} (\omega) {\rm sinc} \left( \frac{\Delta k L}{2} \right) \exp\left(i \frac{\Delta k L}{2}\right) \label{eq:sinc}
\end{eqnarray}
with either
\begin{equation}
P_{\rm I} (\omega)  =  \chi^{(3)}(\omega_{\rm a},\omega_{\rm a},\omega - 2 \omega_{\rm a}) E^2_{\rm a} E^\ast_1 (2 \omega_{\rm a}-\omega)
\label{eq:pI}
\end{equation}
or
\begin{eqnarray}
P_{\rm II} (\omega) & = & \int_0^\infty \chi^{(3)}(\omega',\omega-\omega'+\omega_{\rm a},-\omega_{\rm a}) E_1 (\omega') \nonumber \\
& & \times E_1(\omega-\omega'+\omega_{\rm a}) E_{\rm a}^\ast {\rm d}\omega'. \label{eq:pII}
\end{eqnarray}
Here $\chi^{(3)}$ is the third-order nonlinear susceptibility of the material, $E_1$ the input field. The ancilla field is treated as a continuous wave, i.e., $E_{\rm a}\propto \exp(i \omega_{\rm a} t + i \varphi_{\rm a})$. The vectorial nature of the electric fields only affects the phase mismatch via $\Delta k$, cf.~Figs.~\ref{fig:scheme}(b,c). For type-II self-diffraction, the phase matching terms and the susceptibility can be collapsed into a kernel function
\begin{eqnarray}
K(\omega,\omega') & = &  \frac{\omega}{n(\omega)} {\rm sinc} \left( \frac{\Delta k L}{2} \right) \exp\left(i \frac{\Delta k L}{2}\right) \nonumber \\ & & \times \chi^{(3)}(\omega',\omega-\omega'+\omega_{\rm a},-\omega_{\rm a}) E_{\rm a}^\ast.
\end{eqnarray}
As the field $E_{\rm a}$ is at constant angular frequency $\omega_a$, it is readily measured and drops out of the spectral dependence of the kernel. For narrowband problems, the spectral dependence of $\chi^{(3)}$ is either neglectable or can be sufficiently accounted for by Miller's rule \cite{Boyd,Miller}
\begin{eqnarray}
\chi^{(3)}(\omega,\omega',-\omega_{\rm a}) & \approx &
\frac{q_{\rm e}^4}{m_{\rm e}^3 \omega_0^6 d^5} \chi^{(1)}(\omega) \chi^{(1)}(\omega') \nonumber \\
& \times &  \chi^{(1)}(\omega_{\rm a}) \chi^{(1)}(\omega\!+\!\omega_{\rm a}\!-\!\omega'), \label{eq:miller}
\end{eqnarray}
where $m_{\rm e}$ and $q_{\rm e}$ are electron mass and charge, respectively. $\omega_0$ is deduced from the bandgap $E_{\rm g}=\hbar \omega_0$. Figure \ref{fig:miller} shows a computation of Eq.~(\ref{eq:miller}) for the case of BaF${}_2$. For broadband pulses, the Kramers-Kronig relation discussed in \cite{Sheik-Bahae,Sheik-Bahae2} provides a more accurate description of the dispersion of $\chi^{(3)}$. This formalism provides the nonlinear refractive index $n_2(\omega)$, which can be translated into a more reliable estimate for $\chi^{(3)}$, namely
\begin{eqnarray}
& & \chi^{(3)}(\omega,\omega',-\omega_{\rm a})  \approx 4 \epsilon_0 c \sqrt{n(\omega) n(\omega') n(\omega_{\rm a}) n(\omega\!+\!\omega_{\rm a}\!-\!\omega'} \nonumber \\
 & & \;\;\;\;\;\; \times  \sqrt[4]{n_2(\omega) n_2(\omega')
 n_2(\omega_{\rm a}) n_2(\omega\!+\!\omega_{\rm a}\!-\!\omega')}, \label{eq:SheikBahae}
\end{eqnarray}
where $\epsilon_0$ is the vacuum dielectric constant, $c$ the speed of light, $n$ the refractive index \cite{nBaF2}, and $n_2$ has been derived using the formalism in \cite{Sheik-Bahae,Sheik-Bahae2}. For comparison, this more advanced estimate for $\chi^{(3)}$ is shown in Fig.~\ref{fig:miller}. Both formalisms result in a remarkably small deviation to measured $n_2$ data, which we have corrected for in Fig.~\ref{fig:miller}. Miller's rule may give rise to some artifact at infrared wavelengts, but appears a useful compromise for the Ti:sapphire wavelength range. Use of Eq.~(\ref{eq:SheikBahae}, in contrast, is more limited on the ultraviolet side of the spectrum. In the following, we have considered the full dispersion characteristics of the kernel function, including the full Kramers-Kronig approach for $\chi^{(3)}$ and the phase matching characteristics derived from the Sellmeier equation of BaF${}_2$ \cite{nBaF2}.

Focusing on the case of type-II self-diffraction, we rewrite the complex-valued fields using only real functions
\begin{equation}
E_{1,2}(\omega)= a_{1,2}(\omega) \exp[i\, \varphi_{1,2}(\omega)].
\end{equation}
We finally end up in an operator equation
\begin{equation}
E_2\,=\,\mathcal{F}(E_1)\,,\label{eq:opeq}
\end{equation}
to which we refer to by the operator $\mathcal{F}$ in form of the self-convolution representation
\begin{equation}
[\mathcal{F}(E)](\omega) = \!\!\int\limits_0^\infty K(\omega,\omega')
E(\omega\!-\!\omega'\!+\!\omega_{\rm a}) E(\omega') {\rm d}\omega'\,\!\!. \label{eq:selfconv}
\end{equation}
Here it is important to understand that the SD SPIDER measurement provides the spectral phase $\varphi_2(\omega)$ of $E_2$ while the respective amplitude $a_2(\omega)$ is unknown and cannot reliably be deduced from the readily measurable amplitude of the fundamental field $a_1(\omega)$. Vice versa, while the experimentally known amplitude $a_1$ enters into the convolution of Eq.~(\ref{eq:selfconv}), the respective phase $\varphi_1(\omega)$ is unknown. In the following, we discuss a formalism to solve the inverse problem \cite{Gerth} of finding a fundamental phase $\varphi_1(\omega)$ that agrees with experimental knowledge on $a_1(\omega)$ and $\varphi_2(\omega)$ and best obeys the equation \begin{eqnarray}
 \varphi_2(\omega) & = &\arg \left[ \, \int\limits_{0}^{\infty}K(\omega,\omega') a_1(\omega\!+\!\omega_{\rm a}\!-\!\omega')  a_1(\omega') \right. \nonumber \\
 & \times & \left.\vphantom{\int\limits_{0}^{\infty}} \exp\left[i\varphi_1(\omega\!+\!\omega_{\rm a}\!-\!\omega')+i\varphi_1(\omega')\right]\de \omega'\,\right],
\label{eq:phase}
\end{eqnarray}
which is a data-adapted version of Eq.~(\ref{eq:opeq}) with the operator $\mathcal{F}$ from Eq.~(\ref{eq:selfconv}).

Self-convolution (autoconvolution) equations like Eq.~(\ref{eq:selfconv}) are typically expressed as nonlinear integral equations of quadratic structure \cite{Flemming}. From a mathematical point of view, such equations are usually considered in infinite dimensional Banach spaces of continuous or power-integrable functions. Moreover, they are ill-posed since they characterize inverse problems \cite{Gorenflo,Scherzer,Buerger_Hofmann}. In our case, small errors or noise in experimentally determining $a_1(\omega)$ and $\varphi_2(\omega)$ may lead to arbitrarily large errors in the function to be recovered, namely $\varphi_1(\omega)$. The general way to overcome the ill-posedness \1{employs} regularization methods \cite{Engl,Schuster}, where stable approximate solutions are obtained by solving well-posed auxiliary problems. For the phase retrieval problem under consideration in this paper, we suggest to exploit a very specific variant of a regularization technique \cite{Buerger_Preprint}.
\vspace{-1em}
\section{Regularization for Phase Retrieval}
\label{sec:retrieval}

%
%

In order to find a stable approximation to the solution $\varphi_1$ in Eq.~(\ref{eq:phase}), we regularize Eq.~(\ref{eq:opeq}) in an adapted manner by using a Tikhonov-like approach, with an appropriate regularization parameter
 $\beta>0$, in combination with a gradient method taking into account that instead of the functions $a_1$ and $\varphi_2$, only noisy data $\hat a_1$ and $\hat \varphi_2$, respectively, are available.
To this end, we minimize the functional
\begin{equation}
\begin{split}
 T(E)=\,&\beta\int\limits_0^\infty \left(\hat a_1(\omega)-|E(\omega)|\right)^2 \de \omega\\
 &+\frac{\int\limits_0^\infty \big| [\mathcal{F}(E)](\omega)-|[\mathcal{F}(E)](\omega)| \exp[i\,\hat \varphi_2(\omega)] \big| ^2 \de \omega}{\int\limits_0^\infty \big| [\mathcal{F}(E)](\omega) \big| ^2 \de \omega},
 \end{split}
\end{equation}
yielding the field $E_1^{\rm min}$ that minimizes $T(E)$. We then employ $E_1^{\rm min}$ as an approximation to the solution $E_1$ of Eq.~(\ref{eq:opeq}).
The first summand is a data misfit term and penalizes the deviations between $|E|$ and $\hat a_1$, while the second summand reacts to deviations in the phase of the complex-valued function $[\mathcal{F}(E)](\omega)$ from the measured phase $\hat \varphi_2$. Here normalization in the second term is often required to suppress an oscillatory behavior of the solutions. Suppression of artifacts arising from the ill-posedness of the problem can the be fine tuned by carefully choosing $\beta\ne 1$ and, in turn, emphasizing either summand in the regularization approach. As we did not observe oscillations in the central part of the spectra, we therefore used $\beta=1$ throughout.

\begin{figure}[tb]
\includegraphics[width=\columnwidth]{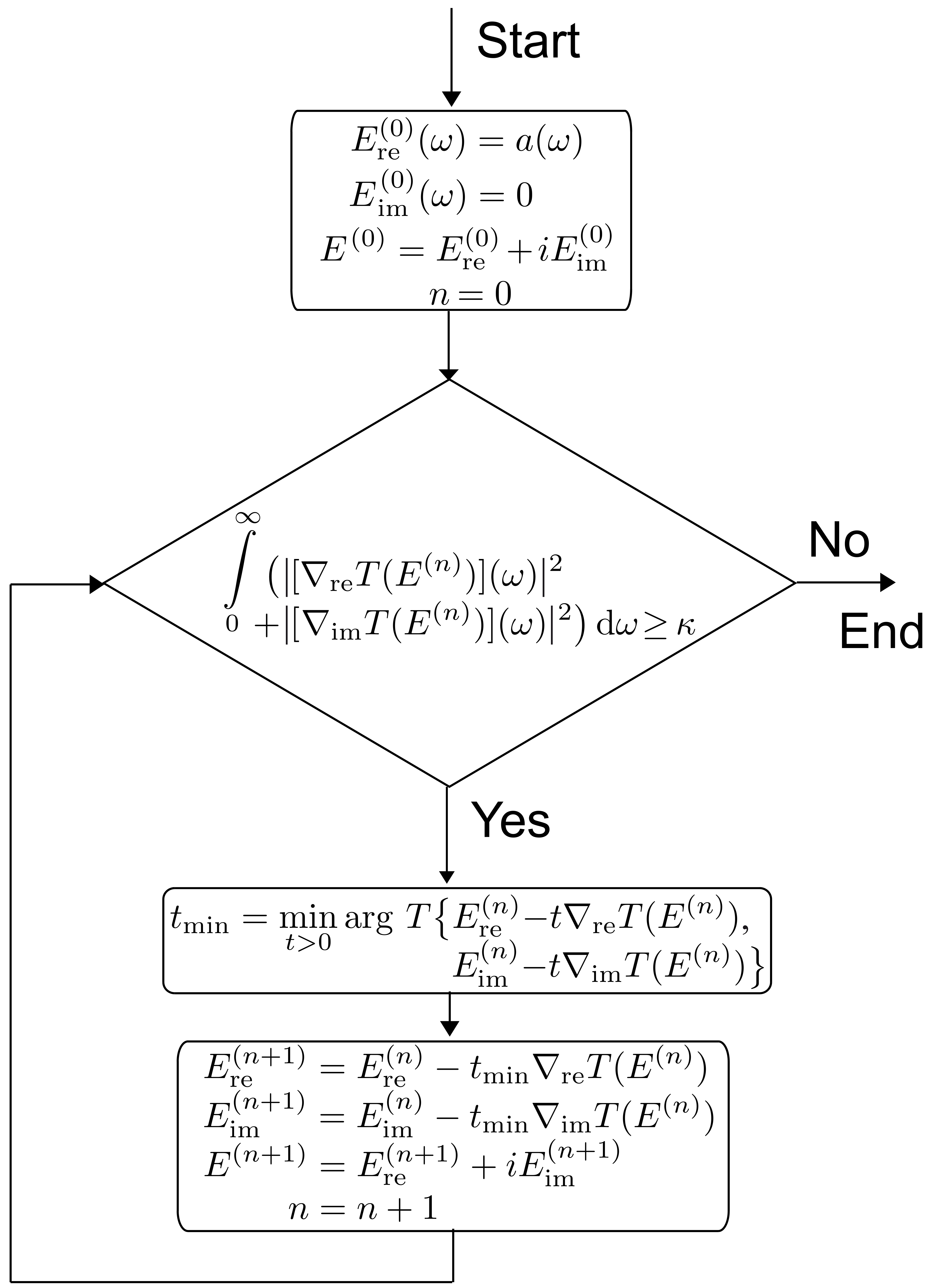}\vspace{-1em}
\footnotesize{\caption{\label{fig:flowchart}Flowchart representation of the regularization algorithm that is used to find the fundamental phase $\varphi_1(\omega)$ that best agrees with measured data and Eq.~(\ref{eq:phase}).}}\vspace{-1em}
\end{figure}

Our goal concerning
 the computational verification of $E_1^{\rm min}$ consists in the construction of an efficient iterative process for finding the minimum of the functional $T(E)$.
Since this functional is not Fr\'echet-differentiable we decompose the electric field $E(\omega)$ into real and imaginary parts, say $E(\omega)=E_{\rm re}(\omega) + i \, E_{\rm im}(\omega)$.
Then we are able to calculate the Fr\'echet derivatives with respect to $E_{\rm re}$ and $E_{\rm im}$ separately. Assuming that $E_{\rm re}$ and $E_{\rm im}$ are square-integrable complex functions over $(0,\infty)$ and using the corresponding norm, this allows us to verify the gradients
$\nabla_{\rm re}T(E)$ and $\nabla_{\rm im}T(E)$.
These definitions enable us to formulate the strategy of our iterative algorithm as follows. While the sum of the norm squares of $\nabla_{\rm re}T(E)$ and $\nabla_{\rm im}T(E)$ is greater than some lower bound $\kappa$, we numerically calculate the optimum step size $t_{\rm min}$ and add the negative of the product of step size and the respective gradient to $E_{\rm re}$ and $E_{\rm im}$. The algorithm is described in detail in Fig.~\ref{fig:flowchart}.

%

For application of the algorithm, it is important to choose the constant $\kappa$ appropriately. If on the one hand $\kappa$ is chosen too large, the algorithm may stop prematurely, i.e., before it is close enough to a solution. If, on the other hand, $\kappa$ is chosen too small, artifacts may arise from the ill-posedness of the underlying problem and may lead to oscillatory behavior of the reconstructed function, as we demonstrate in the following section. For discretization of the problem, we approximate $E_1(\omega)$ and $K(\omega,\omega')$ by piecewise constant basis functions and $E_2(\omega)$ by affine linear functions.
\vspace{-1em}
\section{Demonstration of the Algorithm with Synthetic Data}

To obtain an understanding of the algorithm's reaction to the parameter $\kappa$, let us first test it with synthetic noisy data.
As the exact solution, we choose a function $E_1$ on the frequency domain $[350\text{ THz},410\text{ THz}]$ with Gaussian spectrum $a_1$ normalized to $1$. The spectrum is centered at $380$ THz and covers a $25$ THz full width at half maximum. The spectral phase has been chosen as quadratic.
The kernel function $K(\omega,\omega')$ was chosen identical to experimental data (500\,$\mu$m BaF${}_2$ with 800\,nm ancilla). We interpolate $E_1$ using 1000 points. To the presupposed amplitude $a_1$, we add a relative noise of 1\% yielding the function $\hat a_1$. In a similar fashion, we assume a 5\% additive noise on $E_2$, giving rise to the noise phase $\hat \varphi_2$.
In other words, this means we have noisy data $\hat a_1$ and $\hat \varphi_2$ with $||\hat a_1-a_1|| / ||a_1||=0.01$
and $\hat \varphi_2=\arg \hat E_2$ with $ ||\hat E_2-\mathcal{F}(E_1)|| / ||\mathcal{F}(E_1)||=0.05$.
We set $\beta$ to $1$ and apply the algorithm for different values of $\kappa$.
\begin{figure}[tb]
\includegraphics[width=8cm]{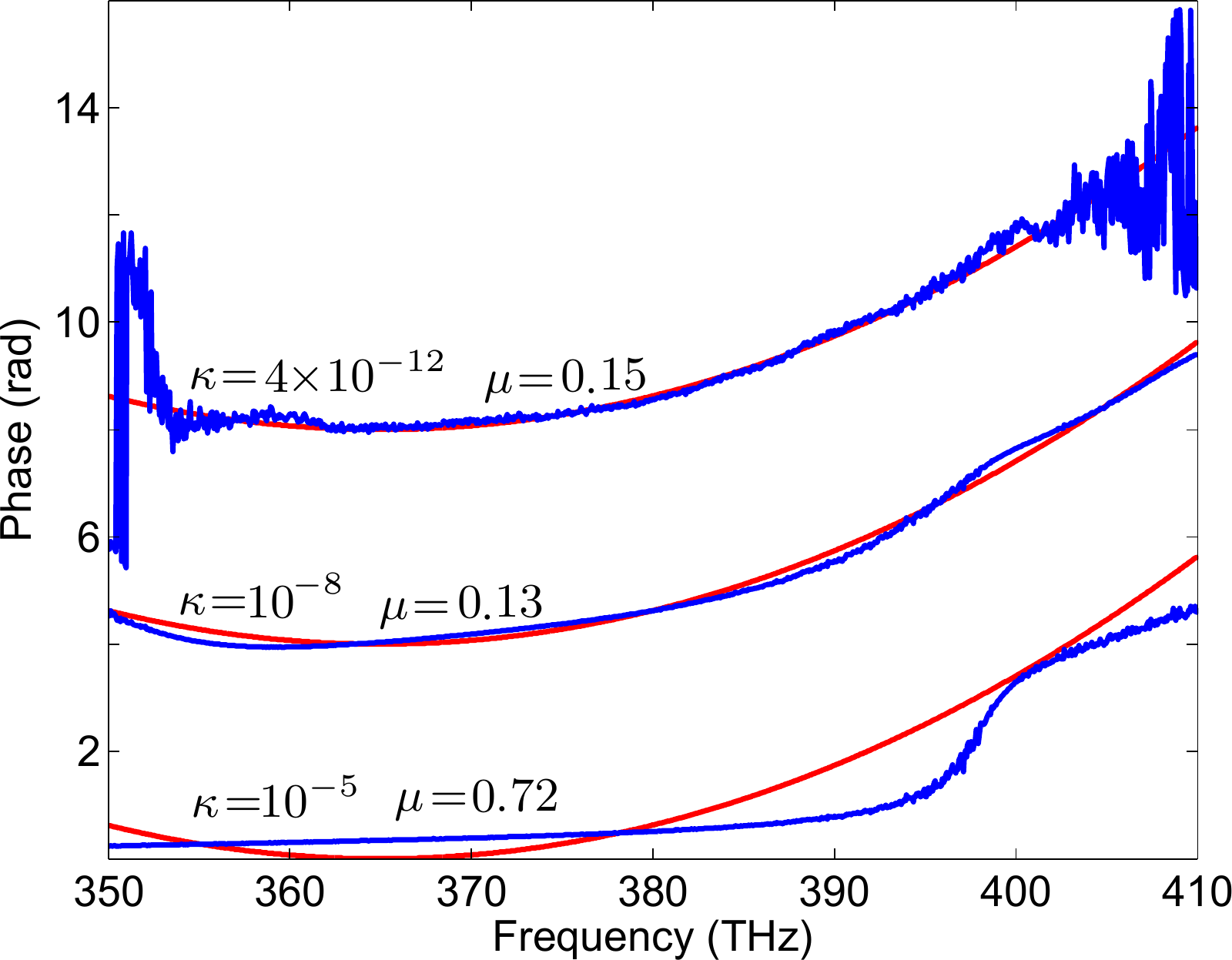}
\vspace{-1em}
\footnotesize{\caption{\label{fig:kappa}Exact phase $\varphi_1$ (red) and reconstructions $\hat \varphi_1$ (blue) for three values of the regularization parameter $\kappa$. For better visibility, the phases have been offset in respect to each other . Reconstructions errors $\mu$ are indicated for comparison.
}}\vspace{-1em}
\end{figure}
As one can see in Fig.~\ref{fig:kappa}, the algorithm stops too early if $\kappa$ is large whereas a very small $\kappa$ leads to oscillations at the boundary. Using $\mu=||\hat E_1-E_1|| / ||E_1||$ to evaluate the quality of the retrieval of $\varphi_1$, we find that values close to $\kappa=10^{-8}$ yield the best overall reconstruction. It should be emphasized that for sufficiently low values of $\kappa\le 10^{-5}$, oscillations of $\varphi_1$ have an effect similar to dispersion oscillations of chirped mirrors \cite{disposci} and neither corrupt reconstructed pulse shape or width, yet give rise to an artificial pedestal structure in the retrieved pulse.
\vspace{-1em}
\section{SD SPIDER Measurements}

In order to demonstrate the experimental feasibility of SD SPIDER, we applied the technique to femtosecond laser pulses from an amplified Ti:sapphire laser \cite{Femtolasers}. The laser system provides pulses with a central wavelength of 800 nm and a pulse energy of 800\,$\mu$J at a repetition rate of 3.2\,kHz. Pulse durations of this laser have been independently verified as 50\,fs. This translates into peak powers of 16\,GW, which can destroy any given optical material when tightly focused. We are therefore limited by the onset of catastrophic optical damage due to six-photon absorption in BaF${}_2$, which is expected to set in at fluences of $\approx 2$ to $3\,$J/cm$^2$ \cite{Mero} at the given pulse duration. In turn, this limits us to intensities below $10^{13}\,$W/cm${}^2$, which is among the highest damage threshold of solid-state dielectric materials. In the experiments, we employed a $500\, \mu$m thick BaF${}_2$ window, which was the thinnest available off the shelf. Given a beam radius $w_0=230\,\mu$m, we estimate that the effective interaction length in the crossed beam geometry is about $450\,\mu$m. Using measured values from \cite{Sheik-Bahae2}, we further estimate $n_2(800\,$nm$)=1.7 \times 10^{-16}$cm$^2/$W, which leads to a maximum expected conversion efficiency on the order of $10^{-4}$ \cite{Agrawal}. This considerations still completely neglects phase matching and the fact that the ancilla is temporally stretched. Taking all these further constraints into account, total conversion efficiencies of $10^{-6}$ or below seem to be more realistic for type II SD SPIDER. For comparison, type I SD SPIDER is expected to yield a 25 times lower efficiency than the type II variant. In conventional SPIDER, one can additionally suppress fundamental stray light by short-pass filters, which is not an option here. Therefore SD SPIDER requires a careful optimization of the crossing angle to keep spurious stray light at an acceptable level. To this end, we found a full external angle $\alpha=3.8^\circ$ to provide sufficient stray light suppression while still providing useful phase matching properties over the entire spectral width of the laser pulses under test.

The experimental implementation of our SD SPIDER setup (Fig.~\ref{fig:setup}) is based on a SFG-SPIDER setup that was originally described in \cite{RevSciInstrum} and has been replicated in commercial SPIDER apparatuses. Here, two replicas of the input laser pulse and the ancilla pulse are generated by reflection off an \'{e}talon and in transmission, respectively. In order to obtain fairly symmetric beam-splitting, the \'{e}talon was oriented at $65^\circ$ and at s-polarization, which provides a nominal $30\%$ Fresnel reflection off a single interface. 
The calibration measurement indicates a fringe period of $0.6$\,THz, which translates into a $1.7\,$ps temporal spacing of the two replicas.
For the induction of a shear $\Omega$ on the latter, a 10\,cm long SF10 glass block with a group-delay dispersion of $15,850\,\rm{fs}^2$ is used. The resulting shear amounts to $\Omega=1.07\times 10^{14}$rad/s. Consequently, the ancilla pulse is chirped to a length of about 1.3 \,ps (FWHM), which closely matches the spacing of the replica pulses. The SD signal is generated by focusing the two beams into a 500\,$\mu$m thick barium fluoride window. As the thickness of this plate may not yet be optimal for wideband spectra, we trestricted ourselves to proof-of-principle demonstration with pulses directly from the amplifier without attempting further compression.

\begin{figure}
\includegraphics[width=8cm]{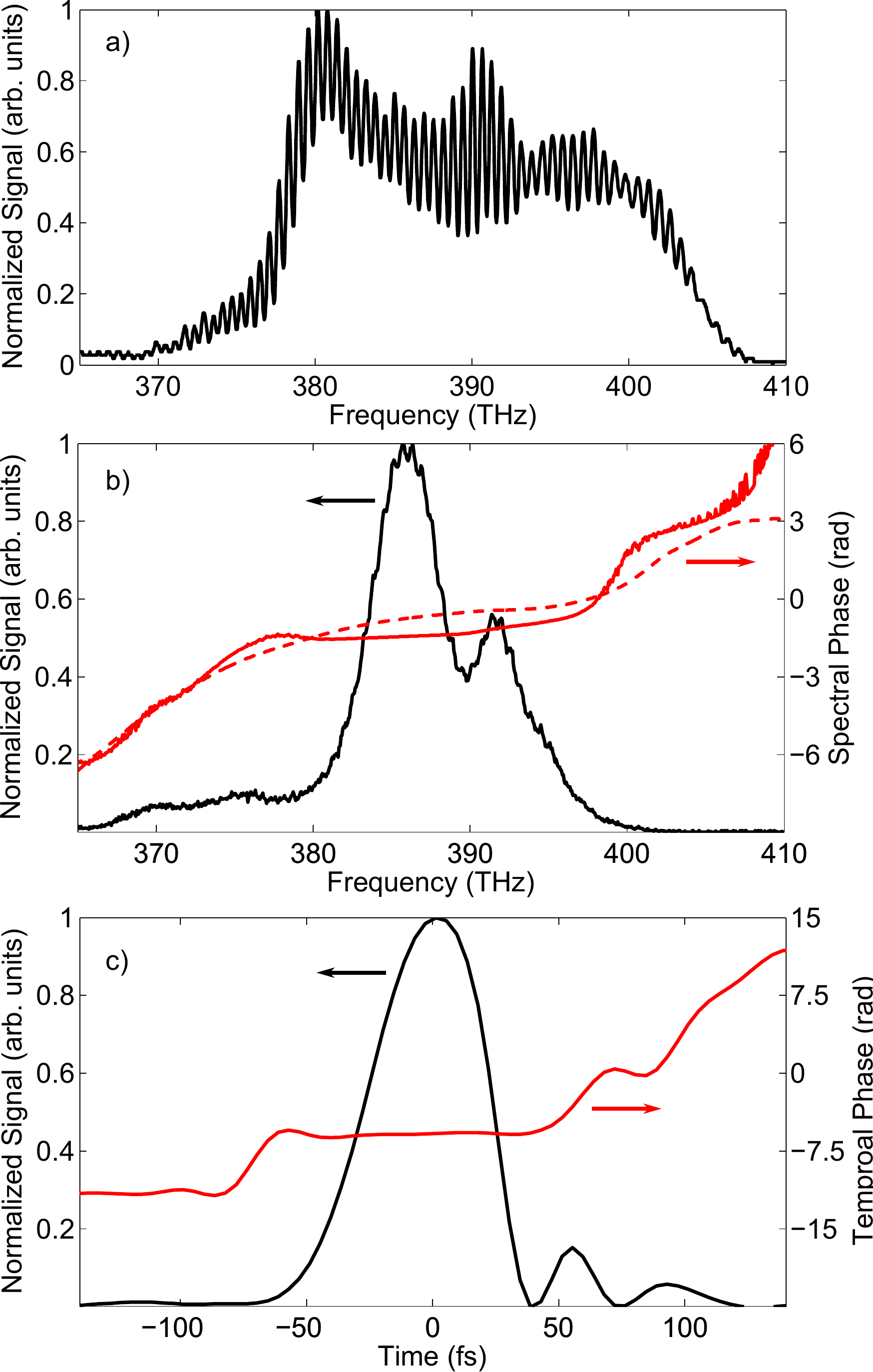}\vspace{-1em}
\caption{\label{fig:exp1} Characterization of a femtosecond laser pulse: a) SD SPIDER trace, b) spectral phase $\varphi_2(\omega)$ of the second harmonic (red dashed), reconstructed spectral phase $\varphi_1(\omega)$ (red) and spectral intensity (black), c) temporal phase (red) and intensity profile (black) of the reconstructed laser pulse.}\vspace{-1em}
\end{figure}

The SD SPIDER trace [Fig.~\ref{fig:exp1}(a)] resulting from interference between the two spectrally sheared SD signals is measured by a 0.5\,m spectrograph with a 300 grooves/mm grating. A 2048 pixel line-scan camera with a pixel size of 13 $\times$ 500 $\mu \rm{m}^2$ captured the the spectograms at an integration time of 5\,ms equivalent to 15 laser shots. The measured interferograms in Fig.~\ref{fig:exp1}(a) contain a decodable interference signal in the range from 370 to 405\,THz, which covers nearly the entire bandwidth of the fundamental laser spectrum in Fig.~\ref{fig:exp1}(b). This spectrum shows a characteristic influence of self-phase modulation, which is being caused by the prism-based compressor \cite{Femtolasers}. The SPIDER signal is clean and strong in the central 15\,THz within the full-width at half maximum of the fundamental spectrum.

Applying the phase reconstruction algorithm of Section \ref{sec:retrieval}, to the discussed SD SPIDER trace yields the spectral phase $\varphi_2(\omega)$ of the second harmonic and the spectral phase $\varphi_1(\omega)$ of the original laser pulse [Fig.~\ref{fig:exp1}(b)]. Figure \ref{fig:exp1}(c) shows the temporal intensity profile and the temporal phase of the reconstructed femtosecond laser pulse. The amplified laser pulses are measured with a duration of 54\,fs, which is within 10\% of the Fourier limited pulse duration of 49\,fs. This reproduces our expectations of a nearly flat phase, which was previously optimized by alignment of the prism compressor directly at the output of the amplifier.
Fitting to the phase curvature in the central part of the trace, we estimate a residual group delay dispersion of 100\,$\rm{fs}^2$, which is equivalent to a 4.5\,m beam path in atmospheric air. As expected, self-phase modulation effects inside the compressor prisms can only be compensated in the central part of the spectrum, where the resulting phase is nearly flat. Beyond this 15\,THz spectral range, the phase quickly rolls off, which causes the observed deviation from the Fourier limit. These observations also agree with measurements of the amplifier output by other independent methods.
\vspace{-1em}
\section{Ultimate Bandwidth}
\begin{figure}
\includegraphics[width=8cm]{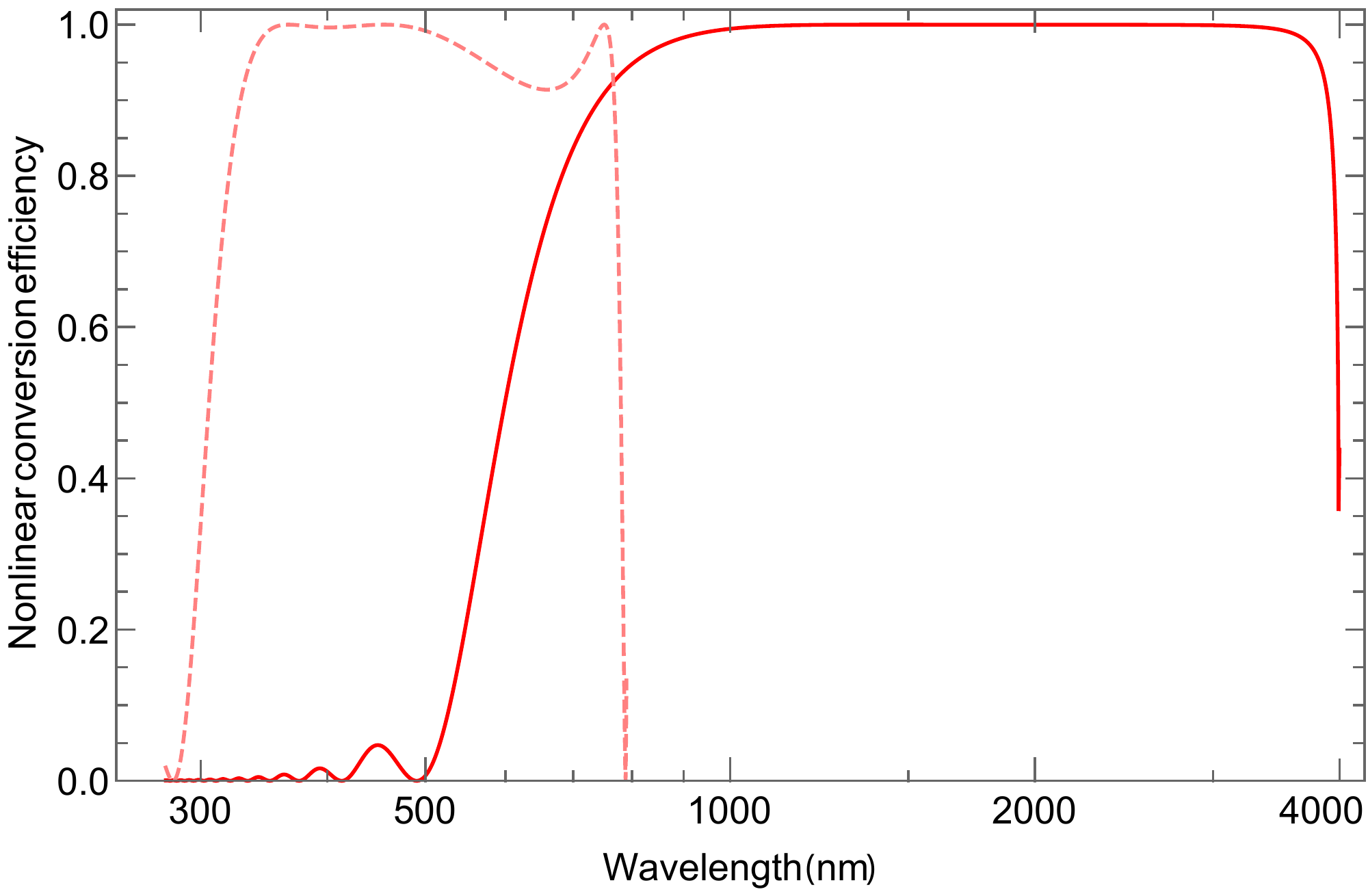}\vspace{-1em}
\footnotesize{\caption{\label{fig:ultimate} Ultimate phase-matching bandwidth of the type-II SD-SPIDER setup, assuming $10\,\mu$m thick BaF${}_2$. $\alpha=3.8^\circ$. Solid curve: ancilla wavelength set to 2150\,nm, 2.3 octave coverage. Dashed curve: ancilla set to 450\,nm, 1.5 octave coverage. The combined conversion bandwidth can host a pulse with a transform limit of 1\,fs.}}\vspace{-1em}
\end{figure}
Our investigation has been limited by the commercial availability of thin BaF${}_2$ substrates. For any applications of self-diffraction, BaF${}_2$ provides highly beneficial properties with its combination of a very high bandgap (9.1\,eV) together with a fairly high $n_2$. Sacrificing some of this advantage, BaF${}_2$ seems to be replaceable by silica, which is available at free-standing thicknesses down to $10\,\mu$m. Another option may be the deposition of a BaF${}_2$ thin film on a carrier substrate.

Let us here speculate that BaF${}_2$ can be obtained at $10\,\mu$m thickness. We first address conversion efficiency considerations. As outlined for our proof-of-principle scenario, we are limited by the onset of catastrophic optical damage. As this phenomenon scales with square root of the pulse duration, e.g., at 4\,fs pulse duration, one expects a $3 \times$ higher damage threshold at $\approx 15$ -- $20 \times10^{12}$W/cm${}^2$. Moreover, scaling to ten times shorter pulse durations enables the reduction of the replica delay to $\approx 100\,$fs and the group delay dispersion of the ancilla to $800\,$fs${}^2$. Notwithstanding manufacturing issues, all these measures together indicate a possible thickness reduction of the BaF${}_2$ sample by a factor 50, i.e., a $10\,\mu$m thickness. Similar consideration can be made for silica as a replacement for ultrathin BaF${}_2$, yet, leading to an overall reduction of the conversion efficiency.

While maintaining the conversion efficiency observed in our experiments, the resulting phase-matching properties are highly favorable for the characterization of single-cycle pulses (Fig.~\ref{fig:ultimate}). Presupposing a $10\,\mu$m BaF${}_2$ substrate thickness, one can establish a 2.3\,octave conversion bandwidth, reaching from 600\,nm to 3\,$\mu$m. This vast bandwidth hosts a $0.6$ cycle pulse at a center wavelength of $1000$\,nm bandwidth, exhibiting a $2$\,fs pulse duration. Adjusting the ancilla accordingly, the scheme is readily adjusted for characterization in the blue, covering the range from 300 to 800\,nm. This corresponds to 1.4 octaves. Combining both, e.g., by a dual exposure SD SPIDER, one can cover a total of 3.3\,octaves. In principle, this detection scheme should be able to detect a single-femtosecond pulse that spans over the entire optical domain from the ultraviolet to the mid-infrared.
\vspace{-1em}
\section{Conclusions}

We presented the foundations of self-diffraction based SPIDER. While $\chi^{(3)}$-based SPIDER variants have been suggested before, these implementations either targeted relatively long pulses as in telecommunication systems or neglected the aspect of retrieving the actual fundamental phase $\varphi_1$ from the measurements. Here we showed how to overcome these obstacles by a numerical regularization strategy. At first sight, it may appear that our strategy combines some disadvantages of SPIDER with those of FROG, namely the one-dimensional support of SPIDER with the necessity of an elaborate retrieval procedure. On the other hand, however, SD SPIDER offers a virtually unlimited bandwidth that can be exploited for the characterization of half-cycle pulses, which seems to be completely out of range for any other characterization technology demonstrated to date. It is certainly clear that selected material aspects, e.g., the material thickness of BaF${}_2$ need to be further developed. Nevertheless, it appears
that SD SPIDER could overcome existing limitations of pulse characterization techniques by at least a factor two, enabling measurement of sub-cycle optical pulses.


\begin{thebibliography}{88}
\bibitem{notegeosmear} Considering an ideal Gaussian beam with a $1/\rm e^2$ beam radius, one expects a far-field half angle $\theta=\lambda/\pi w_0$, with $\lambda$ being the wavelength. Assuming an angular separation of $\alpha = 4 \vartheta$, interference between the two non-collinear beams is estimated to result in spurious modulations below $10^{-3}$. In that case, geometrical delay of $4/\pi$ optical cycles appears at the $1/\rm e^2$ intensity points relative to the center of the nonlinear interaction zone. The rms average beam smearing then amounts to $0.65$ optical cycles, causing a $20\%$ or $5\%$ overestimation of the duration of a Gaussian single-cycle or two-cycle pulse, respectively. Nonideal beam profiles and interference between fundamental and second harmonic can easily double these conservative estimates.
\end{thebibliography}
\end{document}